\documentclass[aps,prb,10pt,twocolumn,superscriptaddress,floatfix,amsmath,amssymb]{revtex4}
\usepackage{boxedminipage}
\usepackage{lscape}
\usepackage{float}
\usepackage[ansinew]{inputenc}
\usepackage{epsfig}
\usepackage{subfigure}
\usepackage{graphicx}
\usepackage{amsmath}
\usepackage{amssymb}
\usepackage{amsthm,url}
\usepackage[french,english]{babel}  
\usepackage{epstopdf}
\usepackage{amsfonts}
\usepackage{color}
\renewcommand{\d}[2]{\frac{#1}{#2}}

\newcommand{\barray}{\begin{eqnarray}}
\newcommand{\earray}{\end{eqnarray}}

\newcommand{\beq}{\begin{equation}}

\newcommand{\eeq}{\end{equation}}

\DeclareMathOperator*{\argmin}{arg\,min}

\newcommand{\G}{\mathbf{G}}
\renewcommand{\d}[2]{\frac{#1}{#2}}

\begin{document}
\selectlanguage{english}

\title{Projected Regression Methods for Inverting Fredholm Integrals: Formalism and Application to Analytical Continuation}
\author{Louis-Fran\c{c}ois Arsenault}
\address{Department of Physics, Columbia University, New York, New York 10027, USA}
\author{Richard Neuberg}
\address{Department of Statistics, Columbia University, New York, New York 10027, USA}
\author{Lauren A. Hannah}
\address{Department of Statistics, Columbia University, New York, New York 10027, USA}
\author{Andrew J. Millis}
\address{Department of Physics, Columbia University, New York, New York 10027, USA}

\date{\today}

\begin{abstract}
We present a machine learning approach to the inversion of Fredholm integrals of the first kind. The approach provides a natural regularization in cases where the inverse of the Fredholm kernel is ill-conditioned. It also provides an efficient and stable treatment of constraints. The key observation is that the stability of the forward problem permits the construction of a large database of outputs for physically meaningful inputs. We apply machine learning to this database to generate a regression function of controlled complexity, which returns approximate solutions for previously unseen inputs; the approximate solutions are then projected onto the subspace of functions satisfying relevant constraints. We also derive and present uncertainty estimates. We illustrate the approach by applying it to  the analytical continuation problem of quantum many-body physics, which involves reconstructing the frequency dependence of physical excitation spectra from data obtained at specific points in the complex frequency plane. Under standard error metrics the method performs as well or better than the Maximum Entropy method for low input noise and is substantially more robust to increased input noise. We expect the methodology to be similarly effective for any problem involving a formally ill-conditioned inversion, provided that the forward problem can be efficiently solved.
\end{abstract}

\maketitle

\hyphenation{Brillouin}

This article is addressed to two different readerships. For physical scientists, we aim to demonstrate, via the solution of a specific and important example, the power of machine learning and statistical regression methods to provide effective solutions to an important class of inverse problems. For statisticians and data scientists we hope to provide an introduction to a class of problems where regression approaches may be applicable.

The class of problems of interest is described by an equation of the form $\mathcal{G}(z)= \mathcal{K}(z,x)\circ \mathcal{A}(x)$ where the kernel $\mathcal{K}$ is known, $\circ$ typically denotes integration, and we are interested in determining $\mathcal{A}$ from measurements of $\mathcal{G}$. This type of problem occurs frequently in the physical sciences. One example is the reconstruction of a potential from  measurements of its effects on incident waves or more generally inversion of a Laplace transform \cite{Epstein08}.  Another example, considered in detail in this paper, is the reconstruction of the values of branch cut discontinuities of a function of a complex variable from data obtained at a sequence of points in the complex plane.  This case is  important in quantum many-body physics, because powerful methods exist \cite{DMFTRev,ClustersRev,DMFTRealMatRev,CTQMCRev} for computing functions along imaginary times or frequencies, while the quantity of direct physical relevance is the spectral function, which describes physical transitions between states and which is related to the magnitude  of a  branch cut discontinuity across the real frequency axis. Inverting the relation to obtain $\mathcal{A}$ given $\mathcal{G}$ is thus important in theory--experiment comparison.

The forward computation of $\mathcal{G}$ given $\mathcal{A}$ typically can be efficiently and stably performed, but reconstructing $\mathcal{A}$  may be difficult because the operators $\mathcal{K}$ typically encountered have many very small eigenvalues, so that the formal inversion
\begin{equation}
\mathcal{A}(x)= \mathcal{K}^{-1}(z,x)  \circ \mathcal{G}(z)
\label{inverse}
\end{equation}
is ill-conditioned: small errors in $\mathcal{G}$ or in carrying out the $\circ$ operation can lead to very large errors in~$\mathcal{A}$. An approximate solution of Eq.~\eqref{inverse} must \textit{regularize} the problem, which means in some way excluding the eigenfunctions corresponding to the very small eigenvalues of $\mathcal{K}$, so that the solution is robust against small errors in~$\mathcal{G}$. Additionally, in many physically relevant cases $\mathcal{A}$ is constrained, for example to be non-negative and to have specified low order moments; it is desirable that approximate solutions of Eq.~(\ref{inverse}) respect such constraints. Finally, an approximate solution for~$\mathcal{A}$ should be accompanied by uncertainty estimates.

In the quantum many-body and materials science communities the standard approach to Eq.~\eqref{inverse} is the Maximum Entropy (MaxEnt) method~\cite{SiSiGu90,Li95}, which approximates  $\mathcal{A}$ as the minimum of an objective function consisting of the sum of the  $\chi^2$ difference between the predicted and actual data for $\mathcal{G}$ and an entropy contribution involving a default model (which incorporates constraints) and a temperature-like parameter that provides a relative weighting of the $\chi^2$ and entropy contributions (a variant, not widely employed due to its computational expense, uses the MaxEnt objective function as the energy in a Monte Carlo process \cite{Sandwick1998,Beach2004}). Experience indicates that certain features can be reasonably well predicted by MaxEnt, but other aspects are the subject of considerable and not easily quantifiable uncertainty \cite{Jarrell96}. The temperature-like parameter is typically chosen phenomenologically. As normally implemented the method does not provide uncertainty estimates (but see [\onlinecite{Sandwick1998,Beach2004}]).

Here, we propose a fundamentally different approach based on modern statistical tools (such methods have recently been applied successfully to solving inverse problems in other areas of the physical sciences, see [\onlinecite{Yevick14,Waller15}]). Our key idea is to treat the solution of Eq.~\eqref{inverse} as a regression problem which is solved by applying statistical machine learning methods to a large database of input--output pairs. The stability of the forward problem means that the needed database of input--output pairs is easily generated, and the formulation as a complexity-constrained regression problem provides a regularization that conditions the formally ill-conditioned inversion. Solutions that respect needed constraints are found by projecting approximated solutions from the unconstrained regression onto the space of functions with the desired properties. An interpolation test is formulated that determines whether a given input $\mathcal{G}$ is sufficiently close to elements of the database that a solution can be trusted. For inputs that pass the interpolation test, we use a quantile regression method to provide pointwise uncertainty estimates.

We demonstrate the power of the approach by applying it to the quantum many-body analytical continuation problem; we call this Analytical Continuation with Projected Regression (ACwPR).  We focus on the case where $\mathcal{G}$ is the electron Green's function, $G$, a correlation function related to the  propagation of an electron in a material at thermal equilibrium at a temperature $T$. The $G$ may be expressed as a function of an imaginary time variable $\tau$ defined on the interval $0\leq \tau\leq \hbar/k_B T$. Powerful numerical methods exist for calculating $G(\tau)$.

In this context, $\mathcal{A}$ is the spectral function $A(\omega)$, a function of a real frequency which has the physical meaning of the density of states for adding or removing an electron at energy $\hbar \omega$ and can be measured in photoemission experiments. Fundamental results of quantum field theory \cite{AGD,Baym1961} imply that $A(\omega)\geq 0$  for all $\omega$, that  $\int \mathrm{d}\omega A(\omega)=1$, and that all higher moments $\int \mathrm{d}\omega\, \omega^lA(\omega)$ ($l\geq 1$) are finite. The integral $\int_{-\infty}^{\infty} \mathrm{d}\omega f(\omega)A(\omega)$ gives the particle density $n$, where $f(\omega)=\left(e^{\hbar\omega/(k_B T)}+1\right)^{-1}$ is the Fermi--Dirac distribution.

The relationship between $G$ and $A$ is \cite{AGD,Baym1961}
\begin{equation}\label{Gbasictau}
	G(\tau) = -\int \mathrm{d}\omega \frac{e^{-\omega\tau}}{1+e^{-\hbar\omega /k_B T}}A(\omega),
\end{equation}
and the problem of inverting Eq.~\eqref{Gbasictau} to infer the quantity of direct physical relevance, $A(\omega)$, is referred to in the physics literature as the analytical continuation problem. The inversion is of the form of Eq.~\eqref{inverse} with $K= \frac{e^{-\omega\tau}}{1+e^{-\hbar\omega / k_B T}}$. The inversion is ill-conditioned, and the complications deriving from the many small eigenvalues of  $\mathcal{K}$ are exacerbated by the fact that the input data typically come from quantum Monte Carlo calculations, which are subject to statistical errors and provide values only at a finite number $\tau$ values. For these reasons, the analytical continuation problem has been a continuing challenge in quantum many-body physics.
\section{Methods}\label{methodssec}
\subsection{Representation of functions}
To apply our machine learning approach, we represent the input $G(\tau)$ and output $A(\omega)$ as vectors. (Henceforth vectors are denoted in boldface and matrices with a tilde overbar. Energy, time and temperature units are chosen such that $\hbar=k_B=1$.)

We represent $G(\tau)$ as $\G$, a vector whose components are the values of $G$ at the points $\tau_p=p /(TP)$ with $p=0, \ldots, P-1$ (many-body theory shows that  $G(0)=1-G(\tau= T))$. In typical many-body calculations a large number of $\tau$ points can be computed; here we use $P=2049$.

Following [\onlinecite{Krivenko2006}] we approximate $A$ as a finite sum of conformal coefficients
\begin{equation}\label{AConfMap1}
	A(\omega) = 2\sum_{m=0}^{L}\text{Re}\{f_mu(\omega)^m\} - f_0,
\end{equation}
where
\begin{equation}
	f_m = \frac{1}{2\pi i}\oint_{|u|=1}\mathrm{d}u\frac{A(\omega(u))}{u^{m+1}},
\end{equation}
with $u(\omega) = (\omega-i\omega_0)/(\omega+i\omega_0)$, $\omega(u)=i\omega_0 (1+u)/(1-u)$ and $\omega_0$ a natural energy unit. This representation is much more compact than a discretization across the region of support of $A$, and it is less susceptible to violations of the non-negativity constraint than the obvious alternative representation in terms of a finite number of orthogonal functions. 
The output is a vector $\mathbf{F}$ of length $L$ which we choose by increasing $L$ until the results cease to change; we find $L=110$ to be sufficient in practice.
\subsection{Databases}
Machine learning approaches require databases of spectral function--Green function pairs. We construct a learning database of $N_{\textrm{learn}}=10000$ examples for training the basic regression, a tuning set of $5000$ examples for determining hyperparameters, and a test set of $5000$ examples for assessing the predictive accuracy. To generate the databases we create spectral functions using simulated data (sums of Gaussians with randomly chosen centers and widths, see supporting information for details) and compute the corresponding Green's functions and conformal coefficients $f_m$ as well as other information including moments and electron densities.

We restrict the $A(\omega)$ that are included to be physically meaningful. We require that $\int \mathrm{d}\omega A(\omega) = 1$ and further constrain the centers and widths such that all $A(\omega)$ have a very small value for $|\omega|$ larger than a cutoff frequency $\omega_{\mathrm{max}}$. We allow high peaks only at small frequency (the physical expectation is that at higher frequencies many decay channels are possible, implying only relatively broad structures in $A(\omega)$ at higher frequencies), and we also impose a smoothness criterion at low frequency that prevents too large a spike in $A$ at any one point. These constraints on $A$ are an important regularization.
\subsection{Unconstrained Regression}
We use a machine learning approach  to perform an unconstrained regression to construct from the learning database a function which provides a predicted output $\mathbf{F}_{\textrm{out}}$ corresponding to an input $\G_{\textrm{in}}$.

Many formulations of regression have been developed and the particular choice of method is not essential. We use kernel ridge regression (KRR) \cite{MLbook}; this  is a  flexible nonparametric regression method which combines the parameter shrinkage properties of ridge regression with kernels to allow for nonlinear relationships. By solving a large number of inverse problems at once, it effectively shares data across problems to reduce relationship uncertainty.

The key ingredients in our KRR are a metric that specifies the distance between two input functions and  a kernel that depends on this distance and controls the size of the region of influence of each example in the database.  We use a simple Gaussian kernel specified by a single hyperparameter $\sigma$,
\begin{equation}\label{kerneleq}
	K(\G, \G^\prime) = \text{e}^{-d^2(\G, \G^\prime)/(2\sigma^2)},
\end{equation}
where $d^2$ is the squared Euclidean distance for continuous functions,
\begin{equation}\label{ddef}
	d^2\left(\G, \G^\prime\right)= \int_0^{1/T}\mathrm{d}\tau\left(G(\tau) - G^\prime(\tau) \right)^2.
\end{equation}
The $m^{\textrm{th}}$ conformal coefficient ($f_{\mathrm{out}}^m$) of the predicted output $\hat{\mathbf{F}}_{\mathrm{out}}$ for a new input example $\G$ is then determined by using the input $\G_{\mathrm{in}}$ to compute  $K_{i}=K(\G_i,\G_{\mathrm{in}})$. $K_i$ is the kernel distance between the input $\G_{\textrm{in}}$ and the $i^{\textrm{th}}$ entry in the learning set. The $K_i$ are assembled into a vector $\mathbf{K}_{\textrm{in}}$ of size $N_{\textrm{learn}}\times 1$. The transpose of this vector is then right-multiplied by an $ N_{\mathrm{learn}}\times 1$ vector $\boldsymbol{\alpha}^m$ to obtain the predicted component $f^m_{\textrm{out}}$:
\begin{equation}
	f^m_{\textrm{out}}=  \mathbf{K}^{\top}_{\mathrm{in}}\boldsymbol{\alpha}^m.
	\label{prediction}
\end{equation}
The unconstrained prediction for the spectral function $\hat{\mathbf{A}}_{\mathrm{out}}$ is obtained by inserting the $f^m_{\mathrm{out}}$ into Eq.~\eqref{AConfMap1}.
The $\boldsymbol{\alpha}^m$ is found by minimizing the regularized squared error loss
\begin{equation}
\sum_i\Big(f^m_i-\sum_jK(i,j)\alpha_j^m\Big)^2+\lambda \sum_{ij}\alpha_iK(i,j)\alpha_j.
\label{Ldef}
\end{equation}
Here  $K(i,j)$ is obtained by using $\G_i$ and $\G_j$ of the learning set in Eq.~\eqref{kerneleq}.
The minimization can be carried out analytically with the result
\begin{equation}\label{ExpCoeffKRR}
	\boldsymbol{\alpha}^m = \left(\tilde{K}+\lambda\tilde{I}_{N_{\mathrm{learn}}}\right)^{-1}  \mathbf{f}^m,
\end{equation}
where $\tilde{K}$ is a $N_{\textrm{learn}}\times N_{\textrm{learn}}$ matrix with entries $K_{ij}$ and $\tilde{I}_{N_{\mathrm{learn}}}$ is the $N_{\mathrm{learn}}\times N_{\mathrm{learn}}$ identity matrix. Because we choose the kernel $K$ and hyperparameters $\sigma$ and $\lambda$ to be independent of $m$, all of the minimizations may be done in parallel.

\begin{figure}
\centering
\includegraphics[width=0.85\columnwidth]{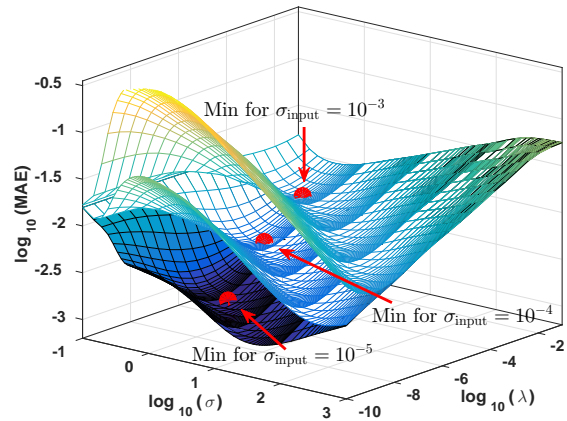}
\caption{(Color online) Mean absolute error (MAE) in the tuning set at different noise level $\sigma_{\mathrm{input}} = 10^{-3}, 10^{-4}$ and $10^{-5}$, as a function of the kernel width $\sigma$ and regularization parameter $\lambda$.}
\label{fig:CV}
\end{figure}

In many physical cases, the input data are noisy, for example when $G(\tau)$ are obtained from quantum Monte Carlo calculations. It is thus important to understand how the predictive accuracy of the model deteriorates as noise is added to $G(\tau)$. While approaches have been proposed for treating inputs with noise (see for example [\onlinecite{Quinonero2003,Girard2005,Mchutchon2011}]), the numerical burden added is substantial. We instead choose the following approach. We train the KRR model parameters using the learning data set without any noise added to $\G$, but we tune the KRR hyperparameters $(\sigma,\lambda)$ using the tuning set with a noisy version of $\G$ as the input for the KRR model. We create the noisy input by adding to each $\G$ in the tuning set a vector of independent mean-zero Gaussian noise with standard deviation $\sigma_{\mathrm{input}}$. We consider three levels for $\sigma_{\mathrm{input}}$, from a fairly large $\sigma_{\mathrm{input}} = 10^{-3}$ to a fairly small $\sigma_{\mathrm{input}} = 10^{-5}$. For each of these noise levels, we select those values for $(\sigma,\lambda)$ that minimize the maximum absolute error (MAE, see the supporting information for details) in the first 10 conformal coefficents $f_m$ when predicting for the tuning set. Fig.~\ref{fig:CV} shows the results. The optimal hyperparameters $(\lambda, \sigma)$ are $(0.574,1.274\times 10^{-3})$ for $\sigma_{\mathrm{input}}=10^{-3}$, $(0.716,1.624\times 10^{-5})$ for $\sigma_{\mathrm{input}}=10^{-4}$ and $(1.483,6.952\times 10^{-8})$ for $\sigma_{\mathrm{input}}=10^{-5}$. While the optimal values of $\sigma$ do not vary much with input noise, $\lambda$ changes by orders of magnitude as the noise is varied. We  also note that the best hyperparameter values will change with database size, composition etc. However, with everything else kept unchanged, the best hyperparameter values are stable with respect to changes in the temperature $T$ at which the $G(\tau)$ were constructed.
\subsection{Projection}
A spectral function obtained from the unconstrained regression Eq.~\eqref{prediction} may violate the constraints of the problem. We therefore correct the unconstrained prediction using a second projection stage that finds the final estimate $\hat{\mathbf{A}}_\mathrm{projected}$ that fulfills the constraints while deviating least from the initial, unconstrained estimate $\hat{\mathbf{A}}$:
\begin{equation}\label{normmin}
	\hat{\mathbf{A}}_\mathrm{projected} = \argmin_{\{\mathbf{A}^*\, :\, \mathbf{A}^* \textrm{ satisfies constraints}\}}  \Vert \mathbf{A}^* - \hat{\mathbf{A}} \Vert_\mathcal{M}.
\end{equation}
Eq.~(\ref{normmin}) is a well-posed convex optimization problem under linear equality and inequality constraints and is easily and efficiently solved (we use a general purpose interior-point quadratic solver). The optimization is formulated directly in terms of the physical spectral function $A(\omega)$ because some constraints (for example, non-negativity) are not easy to express in terms of expansion coefficients such as the conformal amplitudes $f_m$.

In our application, we will include the following constraints:
\begin{itemize}
\item $A$ integrates to one:
$	\int \mathrm{d} \omega A (\omega)   = 1.$
\item $A$ is positive everywhere:
$	A(\omega) \geq 0 \ \forall \ \omega.$
\item The particle density $\int \mathrm{d} \omega (\exp(\omega/T)+1)^{-1} A(\omega)$ for $A$ is known with high precision from $\G$.
\item The first and the second non-central moment of $A$, $	\int \mathrm{d} \omega \, \omega A(\omega)$ and $ \int \mathrm{d} \omega \, \omega^2 A(\omega)$, are also known with high precision from $\G$.
\end{itemize}
The norm  $\Vert \ \Vert_\mathcal{M}$ defines a distance between two functions. It is introduced because the uncertainty in the prediction $\hat{\mathbf{A}}$ is likely not constant across all $\omega$. Furthermore, prediction errors at one location $\omega_a$ may tend to be correlated with prediction errors at a different location $\omega_b$. This suggests that we should require the norm $\Vert \bullet \Vert_\mathcal{M}$ to penalize deviations from $\hat{\mathbf{A}}$ more at locations $\omega_a$ where the predictions are more certain, and that we should take into account the correlation between prediction errors. The squared Mahalanobis distance,
\begin{equation}\label{MahalDist}
	\Vert \bullet \Vert_\mathcal{M} = \bullet^\top \tilde{\Sigma}^{-1} \bullet,
\end{equation}
which weights differences by an appropriate covariance matrix $\Sigma(\omega_a,\omega_b)$ allows us to encode this behavior. We choose the covariance matrix $\tilde{\Sigma}$ in a data-driven way by defining the residual at frequency $\omega$ of a member $j$ of the learning set as
\begin{equation}
R_j(\omega) = A_j(\omega) - \hat A_j(\omega),
\label{residual}
\end{equation}
where $A_j(\omega)$ is the true $A$ and $\hat A_j(\omega)$ is the unconstrained prediction. The average residual at this frequency is $\bar R(\omega) = N_{\mathrm{learn}}^{-1} \sum_{j=1}^{N_{\mathrm{learn}}} R_j(\omega)$. Now, for each pair of frequencies ($\omega_a,\omega_b)$, we define the respective element in the empirical residual covariance matrix as
\begin{equation}\label{EmpCov}
	\Sigma_{ab}^{\mathrm{emp}} = \d{1}{N_{\mathrm{learn}}} \sum_{j=1}^{N_{\mathrm{learn}}} \big(R_j(\omega_a) - \bar R(\omega_a)\big)\big(R_j(\omega_b) - \bar{R}(\omega_b)\big).
\end{equation}
The inverse of the empirical covariance matrix is typically poorly conditioned because the covariance matrix is constructed from a number of pieces of information of the order of the square of the number of frequency points $N_{\omega}$, which will typically be greater than the total number of members of the training set. In these circumstances a principal components estimator  based  on the eigenvalue decomposition $\tilde{\Sigma}^{\mathrm{emp}} = \sum_{k = 1}^{|\{\omega\}|} \lambda_k \mathrm{v}_k \mathrm{v}_k^\top$ is appropriate (see, for example, [\onlinecite{Pourahmadi2013,Bai2011}]). We then retain only the $q$ principal components with the highest eigenvalues, approximating~$\tilde{\Sigma}$ as
\begin{equation}\label{dataCov}
	\hat{\tilde{\Sigma}} = \sum_{k = 1}^{q} \lambda_k \mathrm{v}_k \mathrm{v}_k^\top + \mathrm{diag}\left(\sum_{k = q+1}^{|\{\omega\}|} \lambda_k \mathrm{v}_k \mathrm{v}_k^\top\right).
\end{equation}
The tuning parameter $q$ is chosen to minimize the error between predictions $\hat{\mathbf{A}}_\mathrm{projected}$ and the true densities of states $\mathbf{A}$ on the tuning set and $q=25$ typically suffices in practice. The estimate $\hat{\tilde{\Sigma}}$ is well-conditioned and the inverse, whether taken directly or through an equivalent Cholesky decomposition without direct inversion, has to be performed only once.
\subsection{Assessing Prediction Uncertainty}\label{sec:PredUncert}
Prediction errors occur because finite database size causes imperfect interpolation, and because noise in an input $\G$ will propagate even with perfect prediction.  Quantifying prediction uncertainty is an important part of any solution. We present a two step approach. First we define  an interpolation measure that is used  to determine if an input $\G$ is similar enough to examples in the database so that prediction is appropriate. For inputs $\G$ that pass the interpolation test, we then formulate a quantile regression approach  that provides pointwise uncertainty bounds around the final  predicted $\hat{\mathbf{A}}_\mathrm{projected}$.

\paragraph{An interpolation measure:} To determine whether the input $\G$ is sufficiently similar to examples in the database we analyze the initial unconstrained KRR prediction. An input which is too far away from database samples for the regression to be useful is certainly not appropriate for further analysis. For this purpose, we consider the uncertainty in the predicted conformal coefficients $\hat{\mathbf{F}}$, which is straightforward because KRR is a linear model.

\begin{figure}
\centering
\includegraphics[width=0.8\columnwidth]{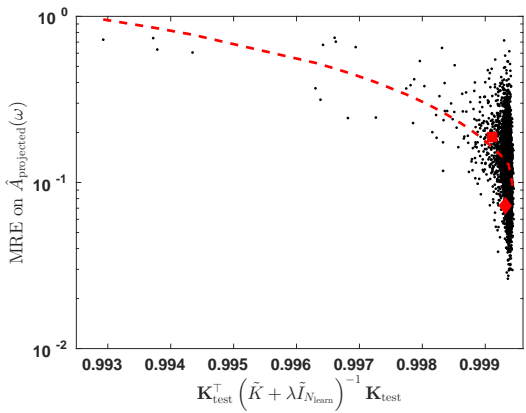}
\caption{(color online) Mean relative error MRE (black dots) and interpolation measure, along with a regression curve (dashed, smoothing spline). Also shown is the location of the examples from Figs.~\ref{fig:A} (red diamond) and \ref{fig:A_indMaxEnt500_490ML_std1En03_predInt} (red square). We see that a good prediction can only be expected if the interpolation measure is very close to one.}
\label{fig:MREvsPostVar}
\end{figure}

Mathematically speaking, we condition the output (the unconstrained vector of conformal coefficients $\hat{\mathbf{F}}$) on the examples in the learning set and the input vector. For a given input $\G_{\mathrm{in}}$ we obtain the vector $\mathbf{K}_{\mathrm{in}}$ as defined in Eq.~\eqref{prediction}. A standard analysis (see, for example, Chap. 2 of [\onlinecite{Rasmussen}]) then says that a measure of confidence in the prediction for our unconstrained vector of conformal coefficients $\hat{\mathbf{F}}$ is
\begin{equation}\label{extrapolationquantity}
	\mathbf{K}_{\mathrm{in}}^{\top} \left(\tilde{K}+\lambda\tilde{I}_{N_{\mathrm{learn}}}\right)^{-1} \mathbf{K}_{\mathrm{in}}.
\end{equation}
This interpolation measure is independent of $m$ because we use the same KRR hyperparameters across all conformal coefficients. A value close to one reflects that the prediction is for an input that is close to many examples in the database, which suggests high prediction accuracy. Decreases from one mean increased prediction error. We  empirically determine a cutoff for deviations from unity below which the prediction quality is typically too low, so that the prediction cannot be trusted. The interpolation metric evaluated for the members of the test set, as well as a measure of the total prediction error on $\hat A_{\mathrm{projected}}(\omega)$, are shown in Fig.~\ref{fig:MREvsPostVar}. We find that in this application the interpolation measure must be within 0.1\% of~1.
\paragraph{Obtaining pointwise prediction bounds:}
If the interpolation metric defined in Eq.~\eqref{extrapolationquantity} is sufficiently close to one,  we proceed to obtain bounds on the prediction error at a given frequency~$\omega$. We treat the residual $R(\omega, \G) = A(\omega, \G) - \hat A_{\mathrm{projected}}(\omega, \G)$ as a random variable with distribution $P_{\omega, \G}$ and study the quantiles of $P$. The quantile $Q^\vartheta(\omega, \G)$ of $R(\omega, \G)$ is the value for which the probability that $R(\omega, \G)\leq Q^\vartheta(\omega, \G)$ is $\vartheta$.
We can therefore write
\begin{align}
	 1 - 2\vartheta &= P( Q^\vartheta(\omega, \G) \leq R(\omega, \G) \leq Q^{1 - \vartheta}(\omega, \G) )\notag \\
    & = P(\hat A_{\mathrm{projected}}(\omega, \G) + Q^\vartheta(\omega, \G) \\ & \leq A(\omega, \G)
    	  \leq \hat A_{\mathrm{projected}}(\omega, \G) + Q^{1 - \vartheta}(\omega, \G)) . \notag
\end{align}
We see that, for $\vartheta < 0.5$,
\begin{equation}\label{predictioninterval}
	[A_{\mathrm{projected}}(\omega, \G) + Q^\vartheta(\omega, \G),\ A_{\mathrm{projected}}(\omega, \G) + Q^{1 - \vartheta}(\omega, \G)]
\end{equation}
is a symmetric, pointwise $1 - 2 \vartheta$ prediction interval for the true, unknown $A(\omega, \G)$. We will focus on pointwise, symmetric 90\% prediction intervals, which means that $\vartheta = 0.05$ in Eq.~\eqref{predictioninterval}.

The quantile functions $Q^\vartheta(\omega, \G)$ and $Q^{1-\vartheta}(\omega, \G)$  are obtained by applying  a quantile regression method~\cite{koenker2005quantile} to a database of $(\G,R)$ pairs. Various quantile regression methods have been discussed; here we adopt the kernel quantile regression method~\cite{takeuchi2006nonparametric} (KQR), a highly flexible quantile regression model that is very similar to KRR. (Note that we use KQR to bound $R$, whereas we use KRR to predict $A$.) Separately for each frequency $\omega$, we model the quantile using KQR as
\begin{equation}
Q(\omega,\G)= \mathbf{K}_{\mathrm{in}}^{\top}\boldsymbol{\beta}(\omega)
\label{Qpredict}
\end{equation}
and estimate $\boldsymbol{\beta}$, separately for each $\omega$, by minimizing the regularized quantile loss
\begin{equation}
	\sum_i\mathcal{L}_{\textrm{quantile}}^\vartheta(R_i,Q_i)+\lambda \sum_{ij}\beta_iK(i,j)\beta_j,
\end{equation}
where
\begin{equation}
	\mathcal{L}_{\textrm{quantile}}^\vartheta(R, Q) =
\begin{cases}
    \vartheta (R - Q),& \text{if } R > Q,\\
    (1-\vartheta)(Q - R),& \text{if } R \leq Q.
\end{cases}
\end{equation}
This minimization needs to be carried out numerically. We use the same distance function, kernel and hyperparameters for KQR as we did for KRR and we reserve more specific training for future work.

Because our approach is based on residuals only, it is agnostic to the underlying model that specifies the relationship between $G(\tau)$ and $A(\omega)$ and can therefore be used with any model that is capable of predicting $A(\omega)$, including MaxEnt.
\begin{figure}[h]
\centering
\includegraphics[width=0.8\columnwidth]{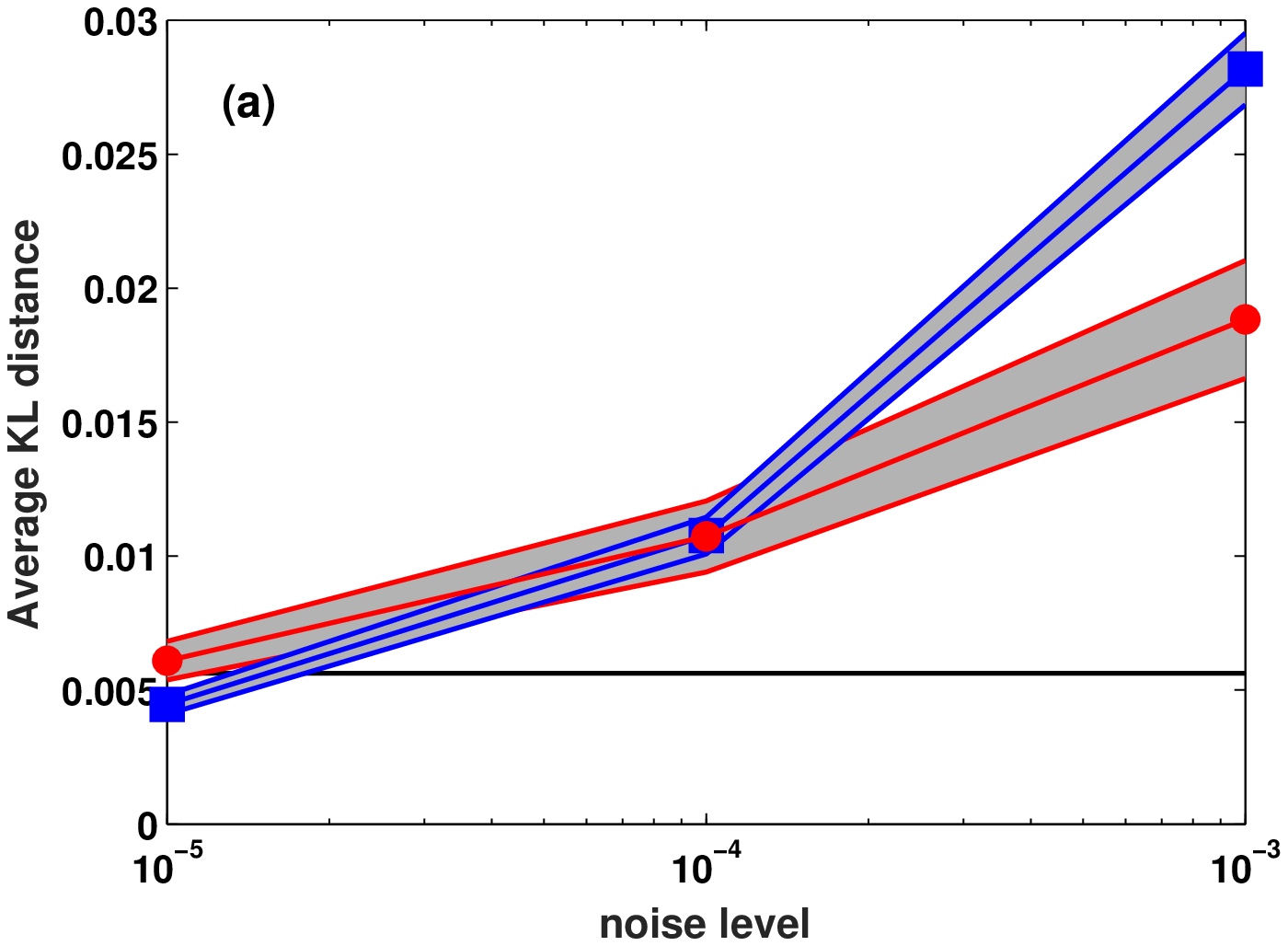}
\includegraphics[width=0.8\columnwidth]{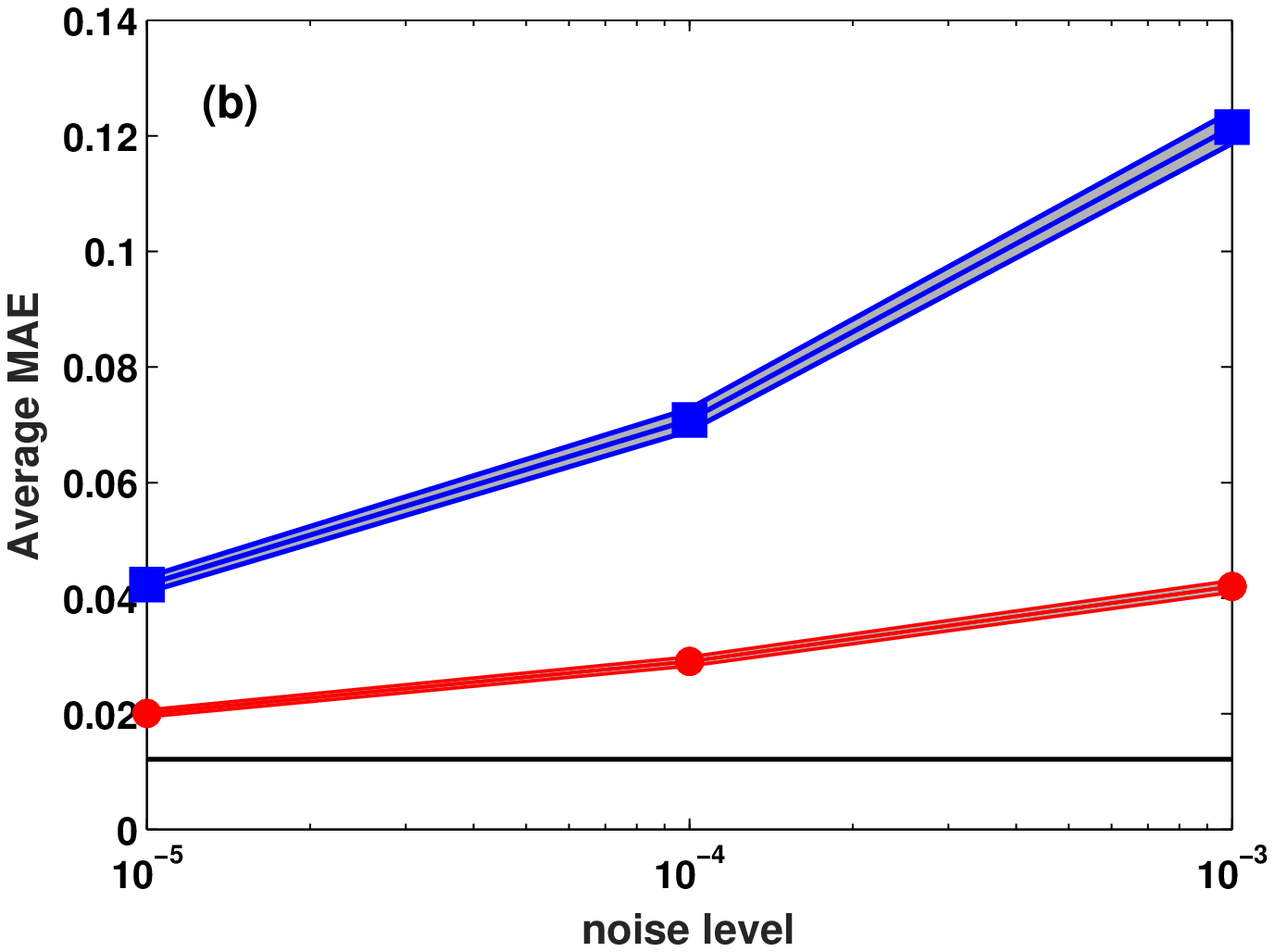}
\caption{(Color online) Average error between predicted and actual spectral function using KL (top) and MAE (bottom) error metrics, averaged over all examples in testing sub-database, for noise levels shown. ACwPR (red line and dots) and MaxEnt (blue line and squares), along with $\pm$ one standard error bounds. The error level obtained using ACwPR for zero noise is shown as a solid black line.}\label{fig:AVGKL}
\end{figure}
\section{Results and Comparison to MaxEnt}\label{sec:illust}
We now assess the performance of ACwPR and compare it to a highly optimized publicly available implementation of MaxEnt, OmegaMaxEnt \cite{Bergeron2015}, using a test data set of 500 observations randomly chosen from the full test set. To simulate the effect of noisy inputs, we perform this comparison at three different levels of delta-correlated noise added to each $G$.

We employ two error metrics: the Kullback--Leibler~(KL) loss
\begin{equation}
	KL_j = \int \mathrm{d}\omega A^j(\omega)\ln\d{A^j(\omega)}{A_{\mathrm{predicted}}^j(\omega)},
\end{equation}
which penalizes relative deviations, and the mean absolute error,
\begin{equation}
	\mathrm{MAE}_j = \frac{1}{N_{\omega}}\sum_{i=1}^{N_{\omega}}\left|A^j(\omega_i)-A_{\mathrm{predicted}}^j(\omega_i)\right|,
\end{equation}
which assesses average deviations between the true function.

Fig.~\ref{fig:AVGKL} shows that according to the KL metric (which MaxEnt is designed to optimize), ACwPR performs minimally worse than MaxEnt at the smallest noise levels on average, but its performance is much better for large noise levels. According to the MAE metric, ACwPR is clearly superior for all input noise levels. Also shown on these figures is the average prediction error computed for noiseless data using ACwPR. We see that the error at the smallest noise level is comparable to the intrinsic prediction error of noiseless data. Comparing ACwPR error and MaxEnt error case by case (see supplemental information) ACwPR performs better in $51\%/67\%/81\%$ of the cases using the KL error metric and $96\%$/$98\%$/$99\%$ of the cases using the MAE error metric at noise levels $\sigma=10^{-5},10^{-4},10^{-3}$.

\begin{figure}[h]
\centering
\includegraphics[width=0.9\columnwidth]{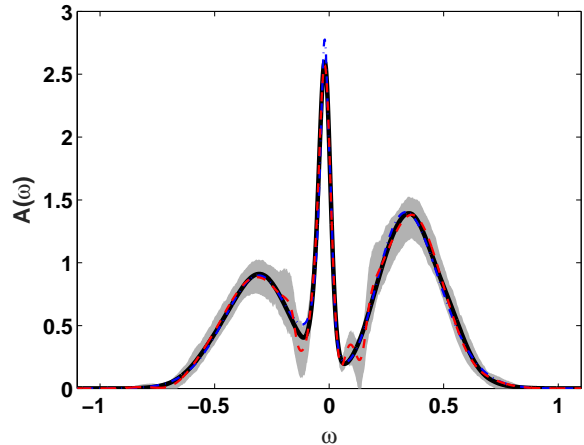}
\caption{True spectral function (black line), projected regression prediction (red $-\,-$ line) and MaxEnt prediction (blue $-\,\cdot$ line) spectral function. Pointwise 5\%--95\% prediction interval for ACwPR in gray.}
\label{fig:A}
\end{figure}

\begin{figure}[h]
\centering
\includegraphics[width=0.9\columnwidth]{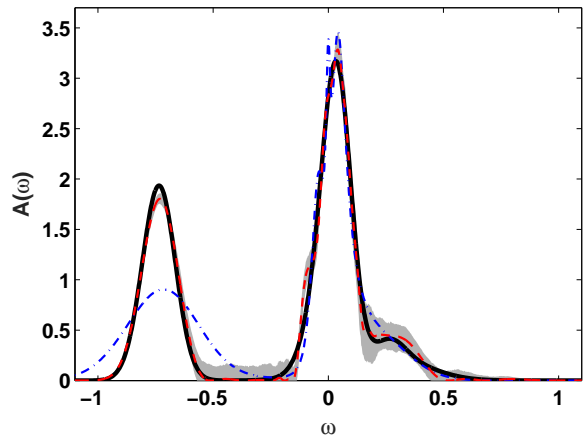}
\caption{True spectral function (black line), ACwPR prediction (red $-\,-$ line) and MaxEnt prediction (blue $-\,\cdot$ line). Pointwise 5\%--95\% prediction interval for ACwPR in gray.}
\label{fig:A_indMaxEnt500_490ML_std1En03_predInt}
\end{figure}

We now have a closer look at two examples at large input noise. We begin with a spectral function characterized by a sharp peak at zero energy and broader structures at high frequency that happens to be the one example with the smallest MAE for MaxEnt. Fig.~\ref{fig:MREvsPostVar} shows that the interpolation metric is very close to one so the prediction can be trusted. We show the true spectral function, the ACwPR prediction with uncertainty bounds and the MaxEnt prediction in Fig.~\ref{fig:A}. For this case, both ACwPR and MaxEnt give good predictions.
Fig.~\ref{fig:A_indMaxEnt500_490ML_std1En03_predInt} shows a more challenging case: a spectral function with a gap (region where $A(\omega)=0$). Once again, Fig.~\ref{fig:MREvsPostVar} shows that the prediction can be trusted. ACwPR performs well, unlike MaxEnt, whose prediction is significantly broadened into the gap.
\section{Conclusion}\label{sec:conclusion}
The inversion of Fredholm integrals of the first kind is an example of an ill-conditioned problem. Any useful method of solution must provide a regularization\cite{Dienstfrey2001}. Direct construction of the inverse operator is challenging and typically not robust to noise in the input data \cite{Epstein08}. In this paper we propose to replace inversion of an operator by regression from a library of known solutions.  Constraints are straightforwardly included by projecting the results of an unconstrained regression onto the subspace of functions satisfying the constraints. Estimates of prediction uncertainty are readily computed from the distribution of the residuals in out-of-sample prediction.

The approach is advantageous for ill-conditioned problems because it provides the needed regularization in three ways: via choice of entries in the database of solved problems, via  representation of the kernel regression in terms of a restricted set of basis functions, and via the tuning parameters used in the kernel. Because the entries in the database of solved problems can be freely chosen, a physical motivated regularization may be employed.

We have applied the approach to the problem of determining a spectral function (density of states of allowed excitations; mathematically, the value of a branch-cut discontinuity of an otherwise analytic function) from measurements of a correlation function. This problem is of considerable importance in the context of condensed matter physics and materials science, and is also representative of a broad class of important problems. Although no particular effort has been invested in optimization, the method was found to compare very favorably to a state of the art and highly optimized maximum entropy implementation. We showed that our approach provides robust reconstruction of spectral functions even in the presence of noisy data and that it captures gap edge behavior well. The approach can be applied in many other similar contexts; the main requirement is that the forward computation is cheap and stable.
\begin{acknowledgments}
L.-F.A. and A.J.M. were supported by the Office of Science of the U.S. Department of Energy under Subcontract No. 3F-3138. R.N. and L.A.H. also received support from Columbia University IDS-ROADS project, UR009033-05. L.-F.A. thanks Dominic Bergeron for support with his software OmegaMaxEnt and Z.~He for insightful discussions.
\end{acknowledgments}


\setcounter{equation}{0}
\setcounter{figure}{0}
\setcounter{table}{0}
\makeatletter

\renewcommand{\theequation}{A\arabic{equation}}
\renewcommand{\thefigure}{A\arabic{figure}}
\renewcommand{\thetable}{A\arabic{table}}
\begin{center}
\textbf{\LARGE Supporting Information}

\textbf{\large Arsenault et al.}
\end{center}

\section*{Database Generation}
\begin{figure}[h!]
\centering
\includegraphics[width=0.8\columnwidth]{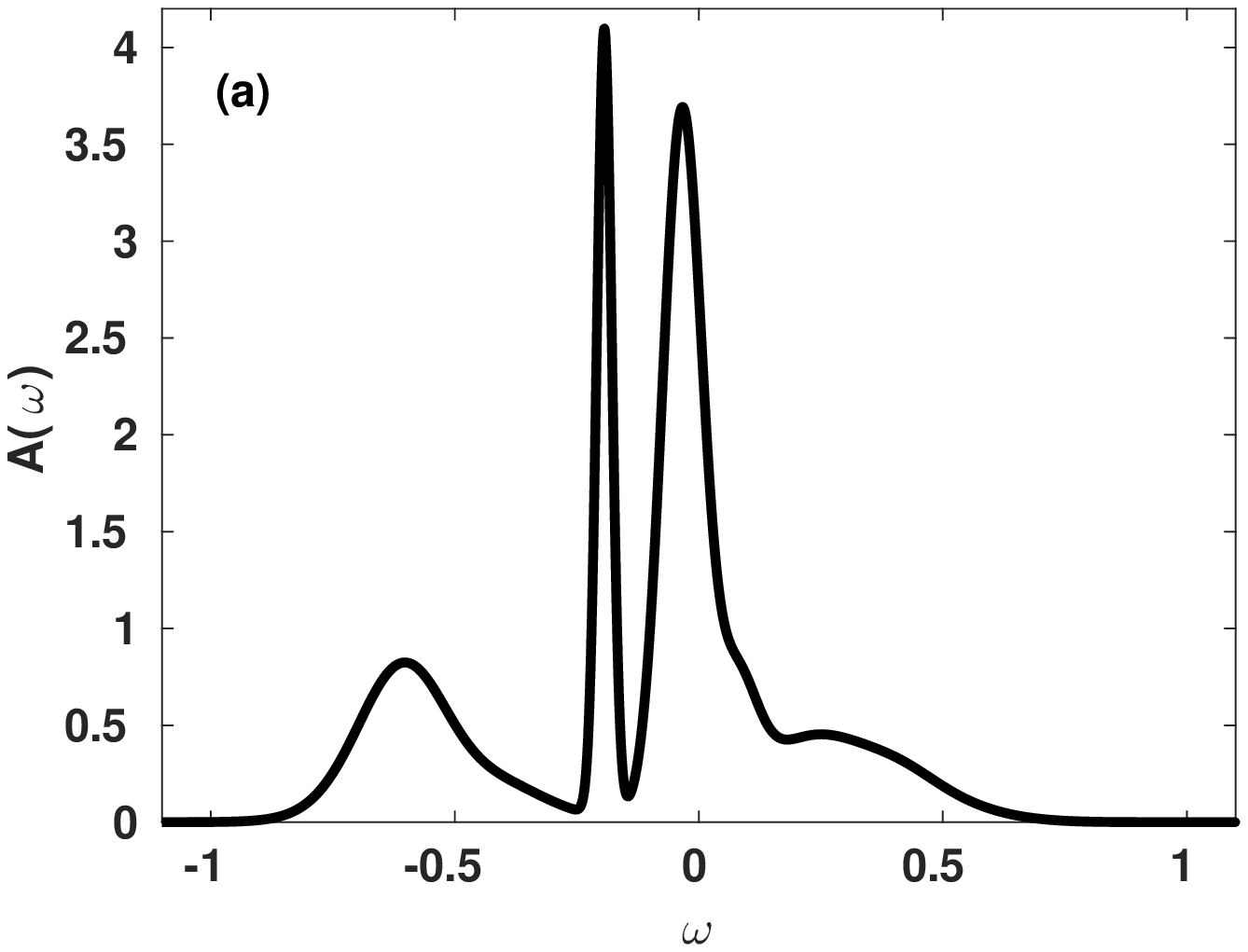}
\includegraphics[width=0.8\columnwidth]{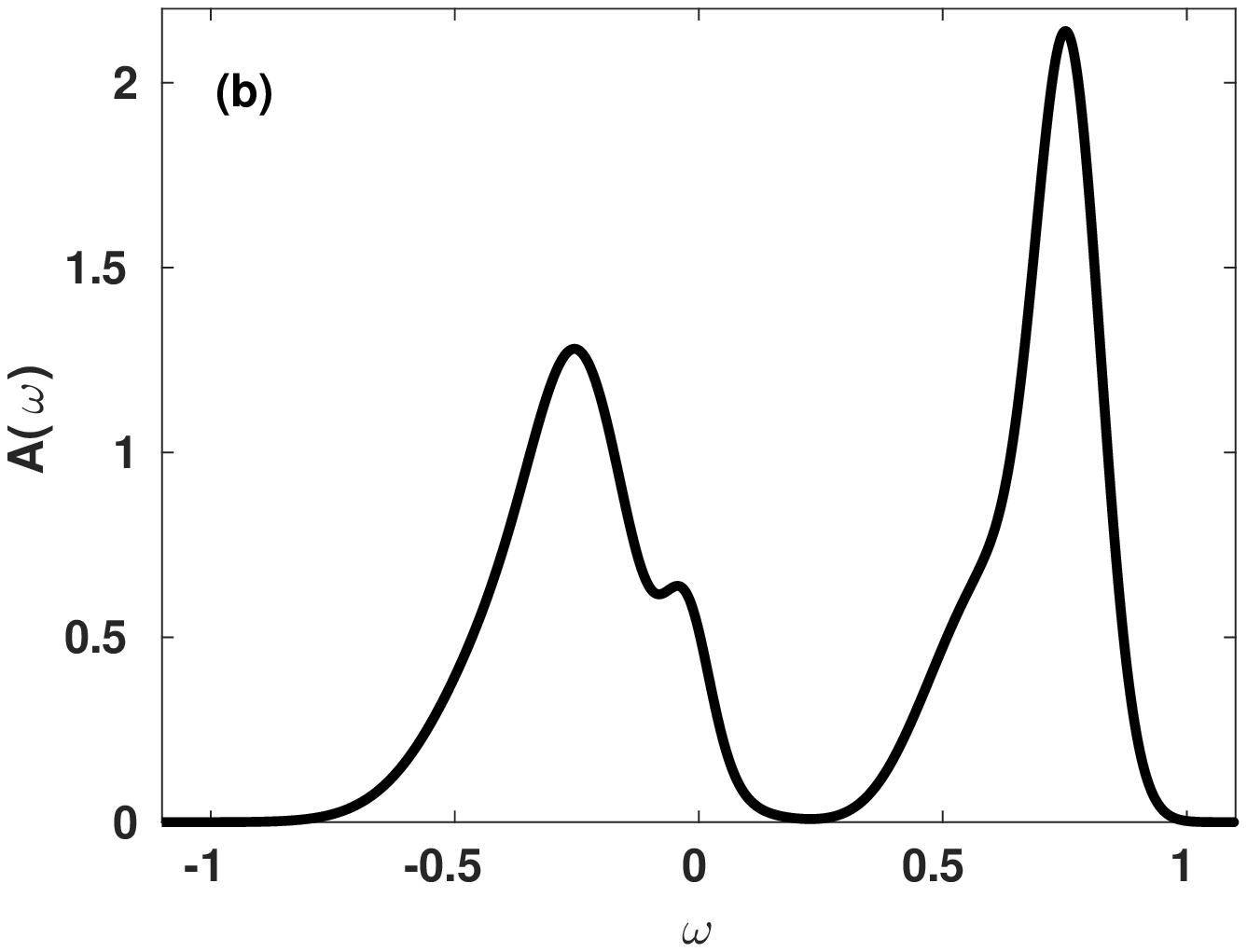}
\caption{Examples of spectral functions $A(\omega)$ in the database}
\label{fig:Aexamples}
\end{figure}

We generate a database of 25,000 $(A(\omega),G(\tau))$ pairs by first generating $A$ and then computing from it $G(\tau)$. Each  $A(\omega)$ in the database is represented as a sum of $R$ Gaussians with  center $\omega_r$, width $\sigma_r$ and weight $a_r$ chosen randomly subject to constraints described below:
\begin{equation}\label{AGau}
	A(\omega) = \sum_{r=1}^R a_r\text{e}^{-(\omega-\omega_r)^2/\sigma_r^2}.
\end{equation}
We use physical insight to restrict the $A(\omega)$ in the database. We imposed that $A$ should have narrow peaks only in a region centered around $\omega=0$. For this purpose, we split the energy window $-1\leq \omega\leq 1$ into three parts using splits at $\omega=-0.2$ and $\omega = 0.2$. We chose different values of $R$ in Eq.~\eqref{AGau} in the range $8 \leq R \leq 33$ and generated centers $\omega_r$ for Gaussians where only up to three are allowed to be in the center region. The widths were chosen such that narrow peaks are only allowed in or close to the center region. We then verify that, with the generated $\omega_r$, $\sigma_r$ and $a_r$, the resulting $A(\omega)$ has small enough a value at $|\omega| = 1$ to be essentially zero chosen here to be $5\times 10^{-4}$. We also imposed a constraint that a peak in the center region cannot be extremely high relative to the rest of the density of states. We impose it by considering the smoothness criterion $\int_{\text{center region}}\mathrm{d}\omega |A''(\omega)| \leq \delta$ as well as a not too large a ratio between the maximum value of $A(\omega)$ in the center region and the maximum value of $A(\omega)$ in the other two regions chosen here to be respectively $1\times 10^{-6}$ and 10.9. Examples of $A(\omega)$ randomly chosen from the database are shown in Fig.~\ref{fig:Aexamples}.

We then compute $G(\tau)$ from $A(\omega)$ at temperature $T=0.005$. The very weak temperature dependence of the hyperparameters means that other temperatures can be considered without difficulty.

\section*{Tuning kernel ridge regression}
We determine the optimal values for $(\sigma,\lambda)$ for KRR as those that minimize MAE between the true and predicted first ten $f_m$ (both real and imaginary parts) out of sample.  For this purpose, we train KRR on the (noiseless) learning set and predict on the (noisy) tuning set. The MAE then is
\begin{align}
	\mathrm{MAE} &= \d{1}{5000}\sum_{j=1}^{5000}\Big(\d{1}{10}\sum_{m=0}^9 \Big[ \left|\text{Re}\left\{f^{\mathrm{KRR},j}_m-f_m^j\right\}\right|\\
    & \qquad \qquad \qquad \qquad \qquad + \left|\text{Im}\left\{f^{\mathrm{KRR},j}_m-f_m^j\right\}\right|\Big]\Big). \notag
\end{align}
We separately tune KRR for three levels of noise, from fairly large $\sigma_{\mathrm{input}} = 10^{-3}$ to fairly small $\sigma_{\mathrm{input}} = 10^{-5}$.
\section*{Tuning the Projection Method}\label{sec:Projec}
To use the projection method, the covariance matrix estimator must first be trained, which means choosing the value of $q$ in Eq.~(14). To do so, we use the learning set and the tuning set. We only use a noise level of $\sigma_{\mathrm{input}} = 10^{-3}$ for validation. Using the KRR model, we make a KRR prediction for the ten thousand $A(\omega)$ of the learning set with noise added to the inputs. We then build the empirical covariance matrix using Eq.~(13). For different values of $q$, we first construct a covariance matrix from Eq.~(14) and then use the tuning set, again with noise added to the inputs $G(\tau)$, to predict the projected $\hat{A}_{\text{projected}}(\omega)$ of Eq.~(10) for the five thousand members of tuning set. We then compute a mean absolute error for these five thousand predictions to obtain the error as a function of $q$ as
\begin{equation}\label{MAEq}
\mathrm{MAE}(q) = \frac{1}{5000}\sum_{j=1}^{5000}\left[\frac{1}{N_{\omega}}\sum_{i=1}^{N_{\omega}}\left|\hat{A}_{\mathrm{projected}}^j\left(\omega_i\right)-A^j\left(\omega_i\right)\right|\right].
\end{equation}
We find that values of $q$ in the range of 16 to 30 offer similar error values and we chose to use $q = 25$, no matter the noise level.
\subsection*{Error of ACwPR Versus Error of MaxEnt}\label{appenError}
\begin{figure*}[h]
\includegraphics[width=1\columnwidth]{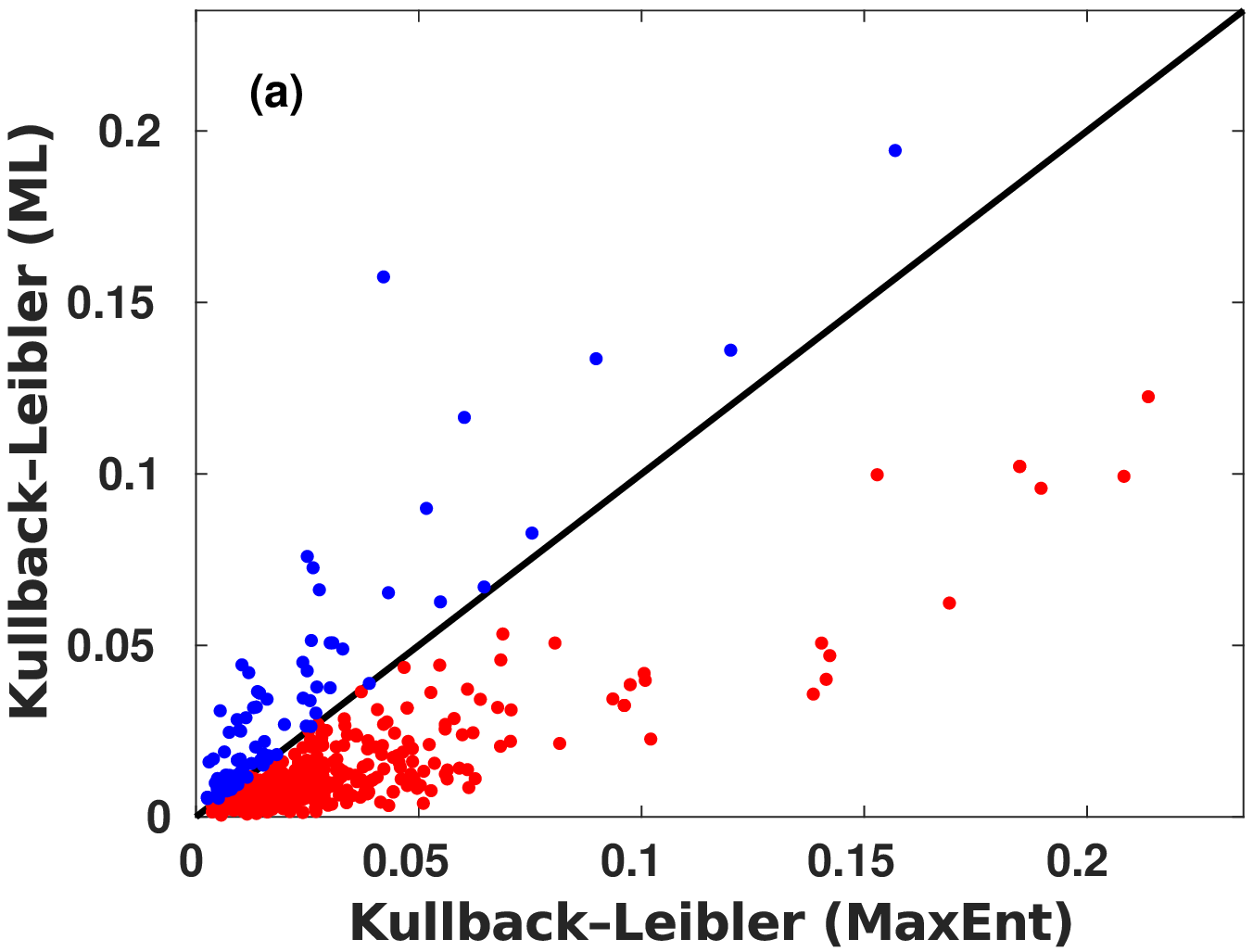}%
\includegraphics[width=1\columnwidth]{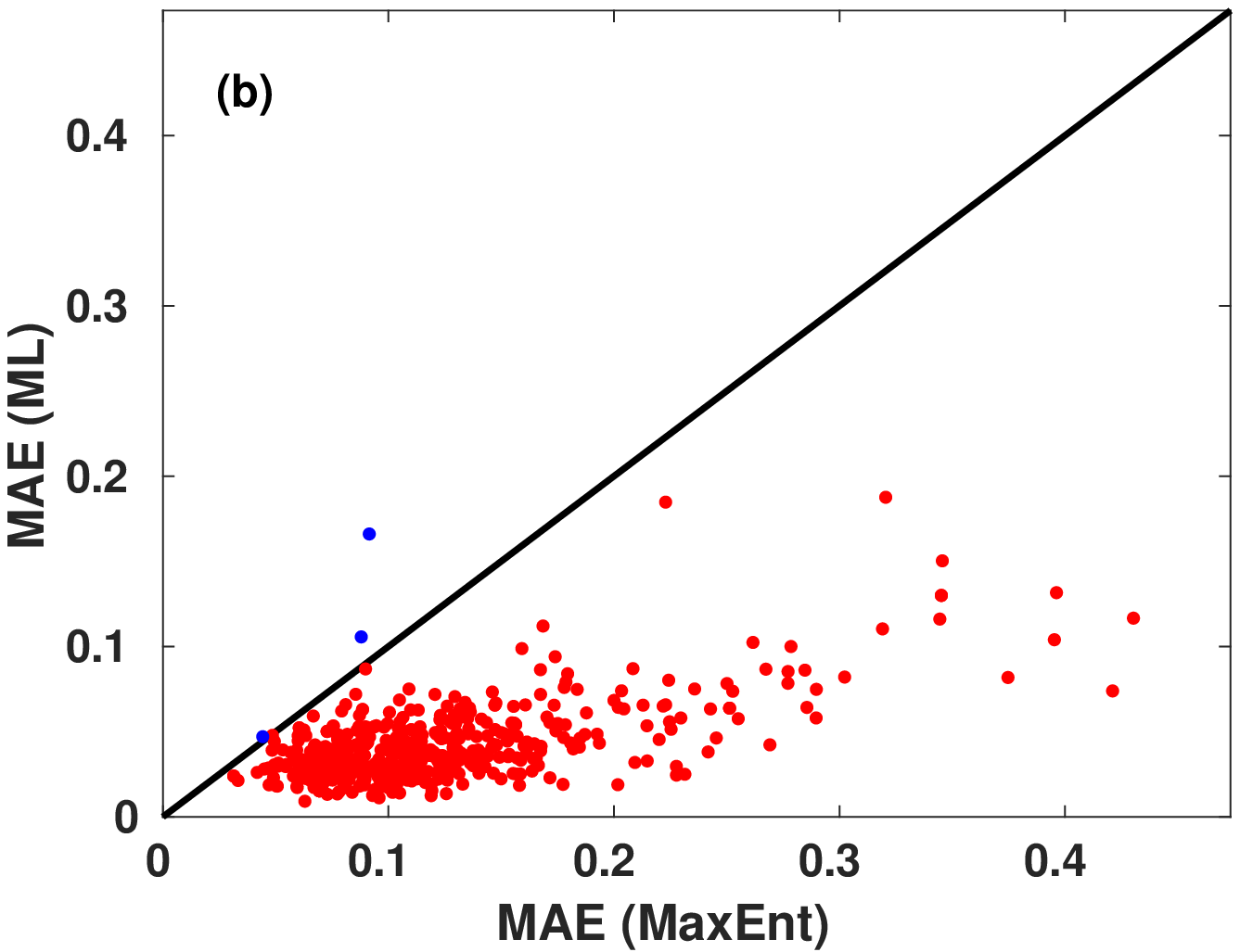}\\
\vspace{0.2in}
\noindent
\includegraphics[width=1\columnwidth]{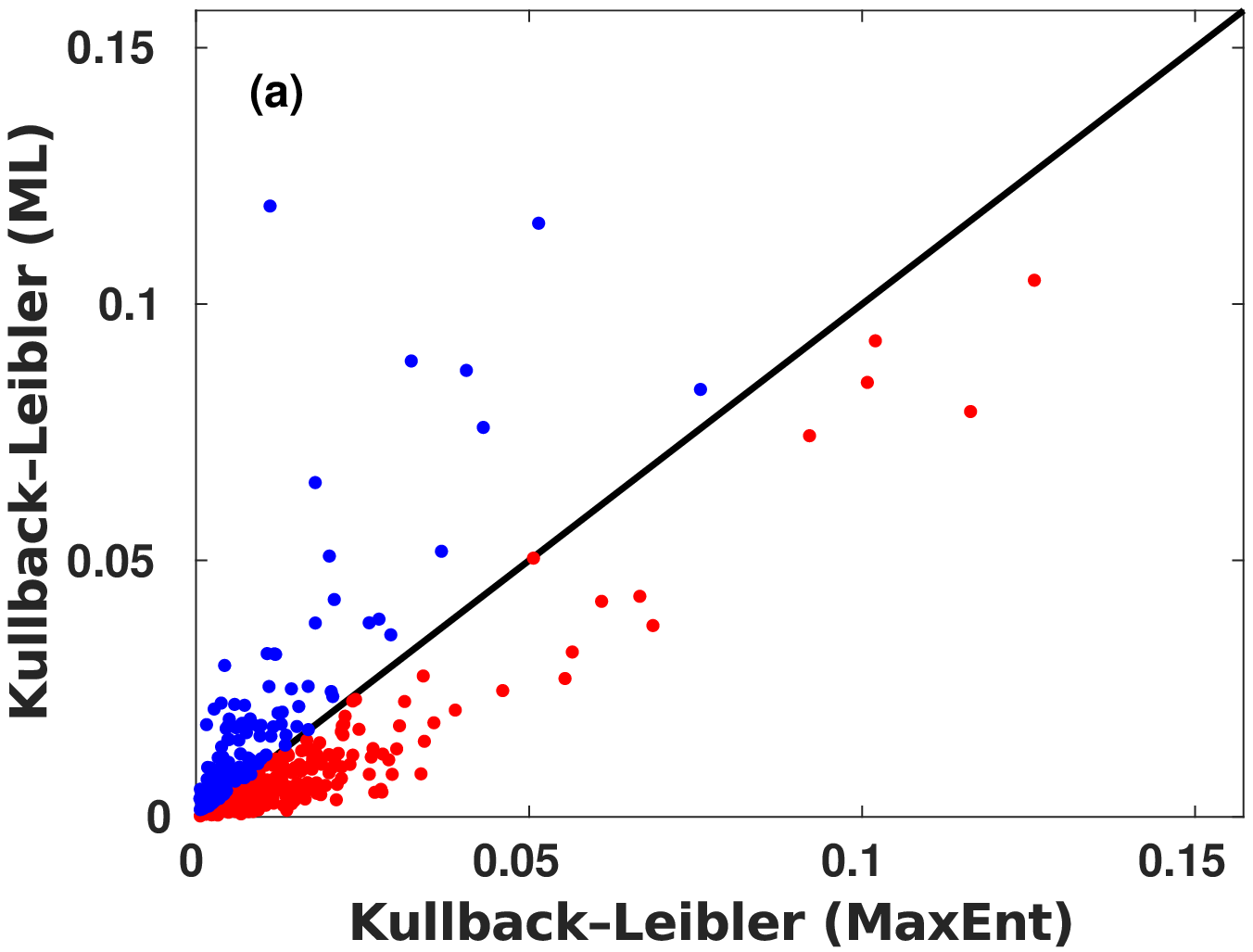}
\includegraphics[width=1\columnwidth]{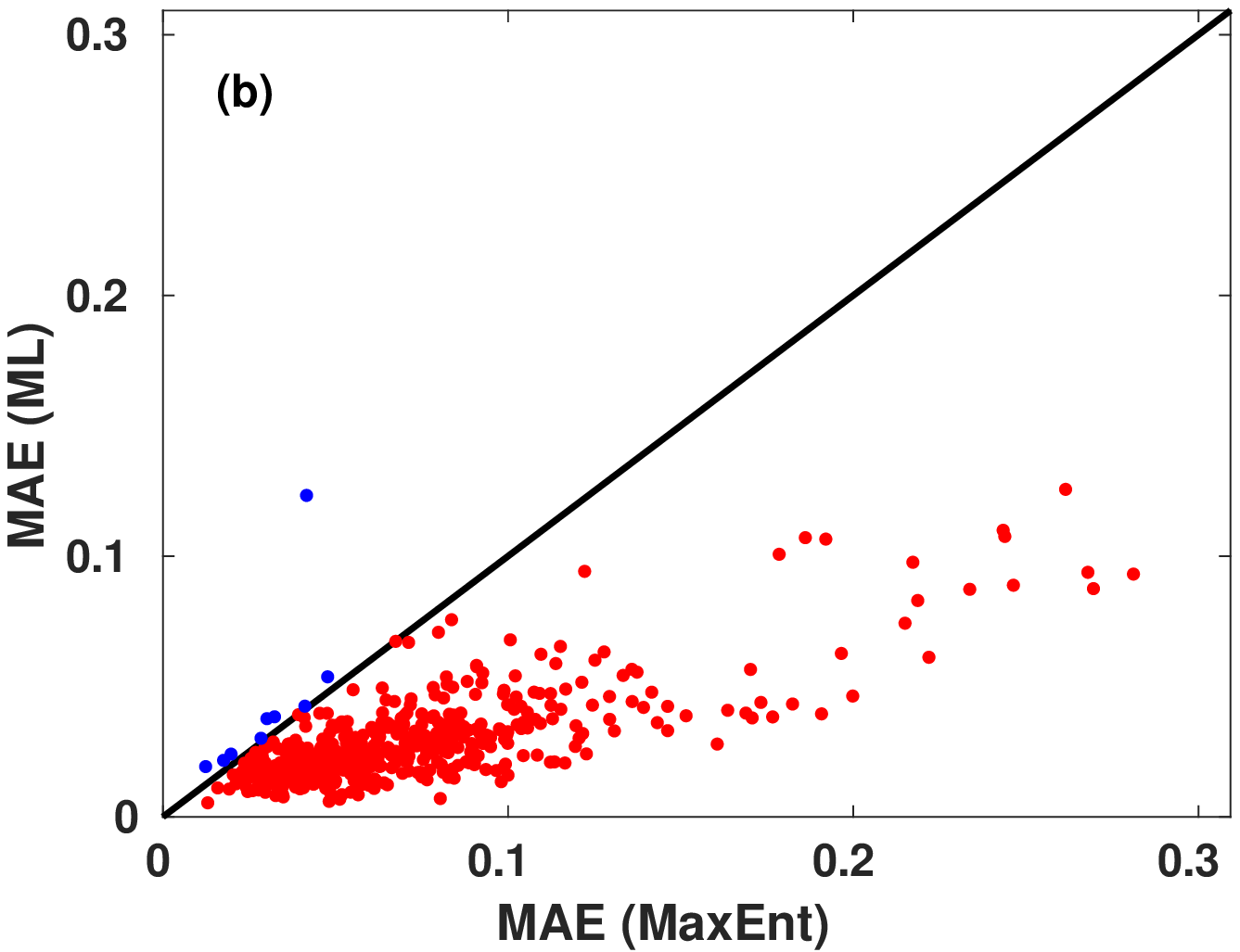}\\
\vspace{0.2in}
\noindent
\includegraphics[width=1\columnwidth]{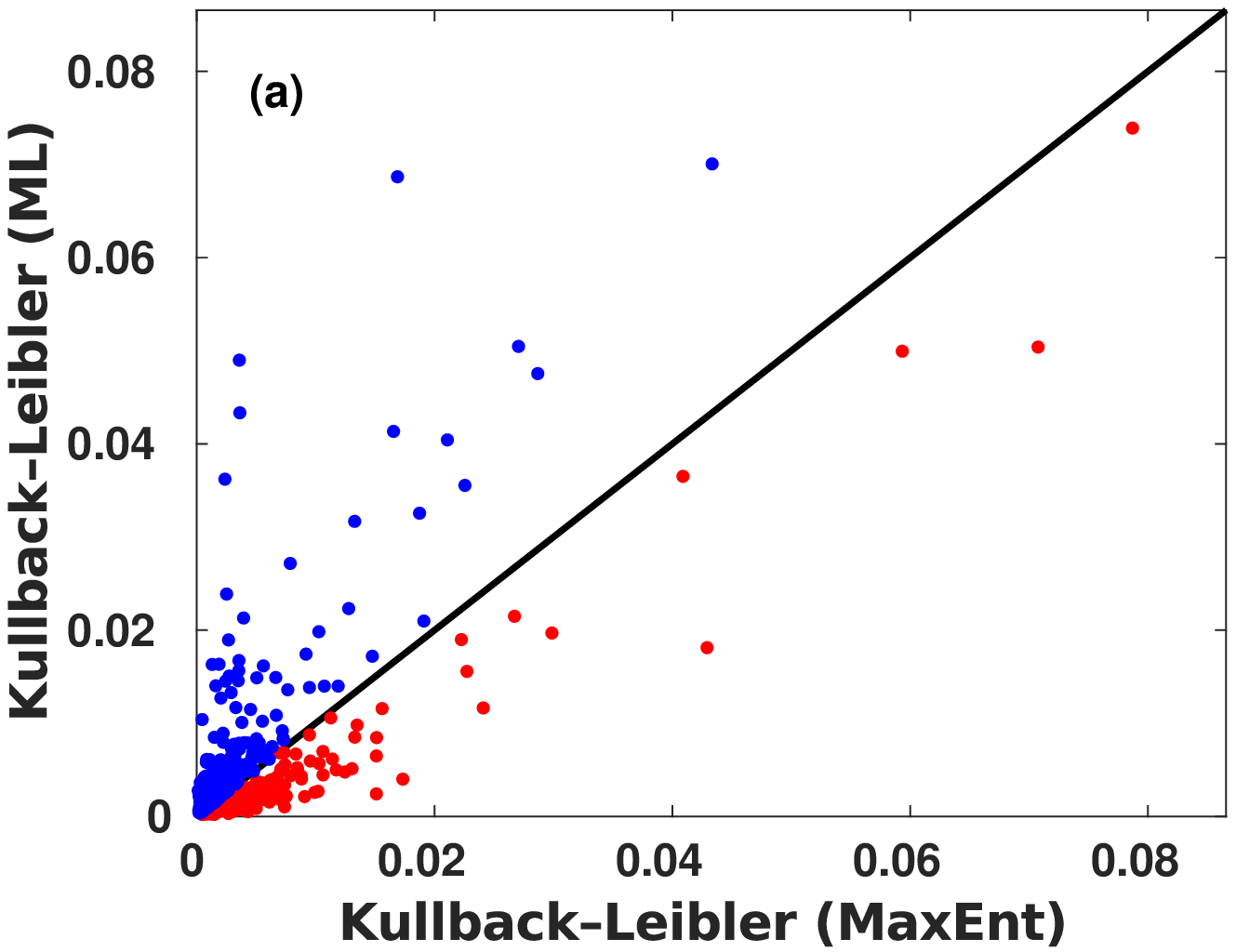}
\includegraphics[width=1\columnwidth]{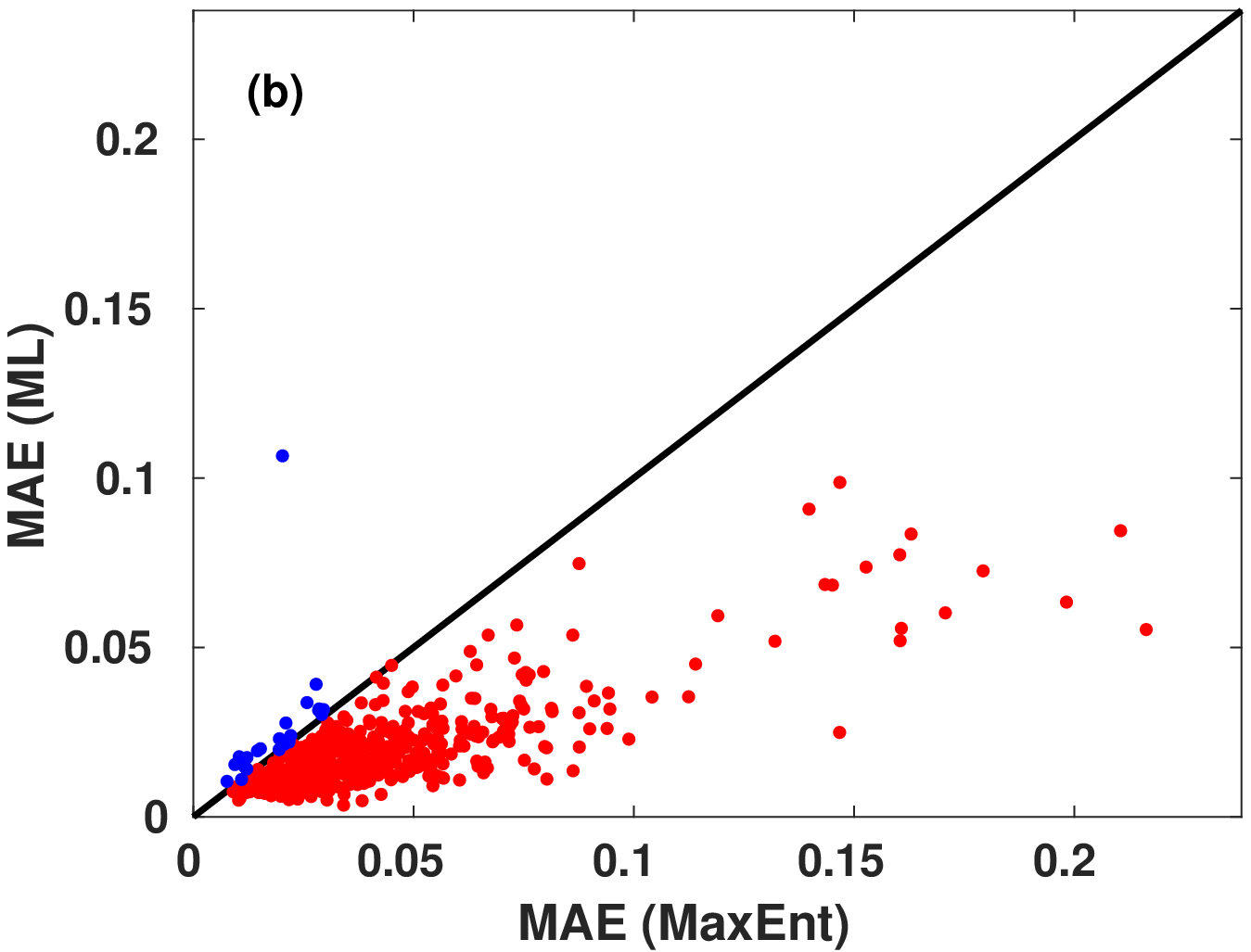}
\caption{(Color online) Kullback-Leibler (left column) and MAE (right column), (a)-(b) errors for noise level $\sigma_{\mathrm{input}}= 10^{-3}$ (top row), (c)-(d) errors for noise level $\sigma_{\mathrm{input}}= 10^{-4}$ (middle row) and (e)-(f) errors for noise level $\sigma_{\mathrm{input}}= 10^{-5}$ (lowest row).}
\label{fig:n03}
\end{figure*}

Fig.~\ref{fig:n03} compares, at three different noise levels, the error of ACwPR to the error of the MaxEnt approach according to both the Kullback--Leibler distance (KL) and the mean absolute error (MAE).  We see that ACwPR performs as well, or better, than MaxEnt at all noise levels.

\end{document}